\documentstyle[aps]{revtex}

\begin{document}
\title{Replica structure of one-dimensional disordered Ising models}
\author{M. Weigt\thanks{martin.weigt@physik.uni-magdeburg.de} \\
      {\small{\it Institut f\"ur Theoretische Physik,
      Otto-von-Guericke-Universit\"at Magdeburg}}\\
     {\small{\it PSF 4120, 39016 Magdeburg, Germany}}\\[.3cm]
      R. Monasson \thanks{monasson@physique.ens.fr}\\
     {\small{\it CNRS-Laboratoire de Physique Th\'eorique de l'Ecole Normale
      Sup\'erieure }}\\
      {\small{\it 24 rue Lhomond, 75231 Paris cedex 05, France}}}
\date{\today }
\maketitle
\begin{abstract}
We analyse the eigenvalue structure of the replicated transfer matrix
of one-dimensional disordered Ising models. In the limit of $n
\rightarrow 0$ replicas, an infinite sequence of transfer matrices is
found, each corresponding to a different irreducible representation
(labelled by a positive integer $\rho$) of the permutation
group. We show that the free energy can be calculated from the replica
symmetric subspace ($\rho =0$).   The other ``replica symmetry broken''
representations ($\rho \ne 0$) are physically meaningful since their
largest eigenvalues $\lambda^{(\rho )}$ control the
disorder--averaged moments $\ll( \langle S_i S_j \rangle - \langle S_i
\rangle \langle S_j \rangle )^\rho \gg \propto (\lambda^{(\rho )})
^{|i-j|}$ of the connected two-points correlations.
\end {abstract}
\vskip .5cm
preprint LPTENS 96/24
\par
PACS numbers : 02.10Sp, 05.50+q, 75.10Nr
\vskip 1cm

In spite of extensive studies during the last twenty years, randomly
disordered systems are not yet fully understood \cite{review}.  One of
the main obstacles to the theoretical analysis of such systems is of
course the lack of translational invariance of the Hamiltonian. The
replica approach, which consists in replacing the original system and
its randomly disordered Hamiltonian by the study of $n$ ($\to 0$)
identical and coupled systems with translationally invariant
interactions, has been proven to be successful in providing a highly
interesting mean--field theory of spin glasses \cite{SK,Par1}.  The
latter relies on a fascinating, but 
mathematically obscur aspect of the replica
mean--field theory, i.e. that the spin glass transition coincides with
the so--called replica symmetry breaking (RSB), or in other words,
with the breaking of the permutation group symmetry of $n\to 0$
elements. In this context, the crucial question is to find the correct
scheme of breaking, allowing an analytical continuation of the theory
when the number $n$ of replicas tends to zero. 
To what extent the physical picture originated in the resolution of
long-range models \cite{mezard} applies to finite-dimensional systems and the
occurrence itself of replica symmetry breaking are still under
question \cite{fisher}.

In this note, we present some preliminary remarks about this
important issue by analysing one--dimensional spin models with 
quenched random couplings
and/or fields.  As in the case of one--dimensional ferromagnetism, no
transition is expected at finite temperature \cite{Luck}. However,
since one--dimensional disordered systems may be solvable to a large
extent with rigorous techniques, they constitute interesting examples
on which the replica symmetry breaking approach can be tested before
being applied to more realistic systems.

To start with, let us consider the Hamiltonian
\begin{equation}
H = - \sum_{i=1}^{N} (J_i S_i S_{i+1} + h_i S_i )
\label{ham}
\end{equation}
with Ising spins $S_i = \pm 1$ and periodic boundary conditions.
The coupling constants and the
external fields are randomly drawn with the probability distribution
$\prod_{i=1}^{N} {\cal P} (J_i) {\it p} (h_i)$.
In general the $2\times2$--transfer matrices
\begin{equation}
T_1 (J_i,h_i) = \left(
\begin{array}{ll}
e^{+\beta J_i + \beta h_i} & e^{-\beta J_i + \beta h_i} \\
e^{-\beta J_i - \beta h_i} & e^{+\beta J_i - \beta h_i}
\end{array}
\right) \label{mat1}
\end{equation}
at temperature $1/\beta$ are not simultaneously diagonalizable. The
calculation of the thermodynamic properties, which requires the
knowledge of the asymptotic properties of the product $\prod_i T_1
(J_i,h_i)$, is rather involved but has been achieved for some
particular choices of disorder distribution, using Dyson's method
\cite{Luck,Derr2,Luck2,bray}.

Hereafter, we shall follow a different route by noticing that, since
the disorder is independently distributed from site to site, the
free energy may be computed through the knowledge of the
replicated and disorder--averaged transfer matrix
$T_n = \ll T_1 (J,h)^{\otimes n} \gg$,
where $\ll (.)\gg$ denotes $\int dJ dh \cdot {\cal P}(J) {\it p}(h)(.)$.
The entries of this $2^n\times 2^n$--matrix are determined by the
replicated spins $\{ S^a =\pm 1, a=1,\ldots ,n\}$.
We introduce the $2^n$ vectors
$ |a_1,a_2,\ldots ,a_\rho \rangle = \bigotimes_a |S^a\rangle$,
$\: a_i \neq a_j \: \forall i \neq j$,
with up spins
$S^a=+1$ at and only at the sites $a \in \{a_1,\ldots ,a_\rho \}$. They
constitute an orthonormal basis of the underlying replicated space $V_n$.
The transfer matrix elements are now given by
\begin{equation}
\langle a_1,\ldots ,a_\rho| T_n |b_1,\ldots ,b_\sigma \rangle = \ll
\exp \left( \beta J \sum_{a=1}^n R^a S^a + \beta h \sum_{a=1}^n R^a
\right) \gg \label{tn}
\end{equation}
with $\{ R^a \}, \, \{ S^b \}$ corresponding to $|a_1,\ldots ,a_\rho
\rangle,\, |b_1,\ldots ,b_\sigma \rangle$. They are obviously
invariant under replica renumbering, i.e. under all transformations of
the permutation group ${\cal S}_n$ given by its $2^n$--dimensional
representation $D( \pi ) |a_1,\ldots ,a_\rho \rangle =
| \pi ( a_1 ),\ldots , \pi (a_\rho ) \rangle$, $\forall \pi \in {\cal S}_n$.
Every irreducible decomposition of this representation $D$ is isomorphous to
the most general eigenspace decomposition of $V_n$.
$D$ already appears in the theory of atomic spectra, for a detailed 
presentation see \cite{Wig}.  The number of up spins in the basis 
vectors of $V_n$ remains clearly invariant under replica
permutations. So we find $n+1$ subrepresentations $\Delta_\rho$
which are carried by the
spaces spanned by $\{ |a_1,\ldots ,a_\rho \rangle,\,1\leq a_1
< \ldots < a_\rho \leq n\}$, $\rho=0,\ldots ,n$.
But for $\rho \neq 0,n$ these $\Delta_\rho$ are
still reducible.  Consider e.g. the vector
$\sum_{a\neq a_i} |a,a_1,\ldots ,
a_{\rho-1} \rangle $; under permutations it behaves like
$|a_1,\ldots ,a_{\rho-1} \rangle $, so we find
$\Delta_\rho \cong \Delta_{\rho-1} \oplus D_\rho$ $\forall \rho \leq n/2$.
Using in addition the
symmetry $\rho \mapsto n-\rho ,\, |\pm \rangle\, \mapsto |\mp \rangle$,
we obtain the complete decomposition of $D$:
$\Delta_0  \cong  D_0$,
$\Delta_1  \cong  D_0 \oplus D_1$,
$\ldots$,
$\Delta_\rho  \cong  D_0 \oplus D_1 \oplus \ldots \oplus
D_{\min(\rho,n-\rho)}$,
$\ldots$,
$\Delta_n  \cong D_0$, whose irreducibility has been proven in \cite{Wig}.
By changing the basis of $V_n$ with respect to this decomposition and taking
at first the vectors of all $D_0$-spaces, then those of all $D_1$-spaces and
so on, we can block-diagonalize the transfer matrix $T_n$. 

Replica symmetry (RS) corresponds to the restriction to the first
block since all $D_0$--vectors are invariant under permutations. 
Due to the representation structure, the RS-transfer matrix has 
$n+1$ non-degenerate eigenvalues and reads,
cf. \cite{pendry}
\begin{equation}
T_n^{(0)} (\sigma,\tau) = \sum_{\mu = \mu_-}^{\mu_+} {\sigma \choose \mu}
 { n-\sigma \choose \tau - \mu }\  \ll
\mbox{exp} \{ \beta J ( n+4\mu -2\tau -2\sigma )
 + \beta h (2\sigma -n) \} \gg
\label{trs}
\end{equation}
where $\mu_- =$ max$( 0, \sigma + \tau -n)$ and $\mu_+ =$ min$( \sigma,\tau )$
and the indices $\sigma ,\tau$ run from $0$ to $n$.
The RS site-dependent partition function is given by the
iterative description
$Z_{i+1}(\tau) = \sum_{\sigma=0}^{n} T_n^{(0)}( \sigma,\tau ) Z_i (\sigma)$.
Introducing the generating function
$Z_i[x] = \sum_\sigma Z_i(\sigma) x^\sigma$, the latter reads
\begin{equation}
Z_{i+1}[x] = \int _0 ^{\infty} dy \ll e^{-\beta h n} 
( e^{\beta J}+ x e^{-\beta J} )^n \delta\left( y -f(x)\right) \gg Z_i[y]
\label{mat}
\end{equation}
where 
\begin{equation}
f(x)=e^{2\beta h} \frac{ e^{-\beta J} + x e^{+\beta J}}{ e^{+\beta J} + x
e^{-\beta J}}\;. \label{f}
\end{equation}
In the thermodynamic limit, we call $\Phi(x)$ the right eigenfunction 
of $T_{n\to 0}^{(0)}$ which has the (maximal) eigenvalue unity and an integral
normalized to one,
\begin{equation} 
\Phi(x) =  \int_0^\infty dy \,\ll \delta ( x-f(y) ) \gg\Phi(y)\;.
\label{density}
\end{equation}
The free energy density is given by the O($n$)-corrections in (\ref{mat}),
\begin{equation}
f = \ \ll h \gg - \frac{1}{\beta}\int_0^\infty dx \, \ll 
\log(e^{\beta J}+xe^{-\beta J}) \gg \Phi(x) \,
\label{freeenergy}
\end{equation}
Equations (\ref{density}) and (\ref{freeenergy}) for the random field Ising
model are precisely the results derived in \cite{Derr} by analysing
the Lyapunov exponent of the infinite product of disordered transfer
matrices $T_1(J_i,h_i)$ given in (\ref{mat1}). RS gives
therefore the correct result for the free energy, and
(\ref{density}) is usually interpreted as the Dyson--Schmidt
equation for the invariant density $\Phi$ of the local fields \cite{Luck}.
A similar result was already established by Lin \cite{Lin} who showed 
the equivalence of an early replica method developped by Kac with Dyson's 
approach for the phonon spectrum of a chain of random masses and springs.

To be sure that replica symmetry is not violated, we have to check
that the eigenvalue unity is not degenerate, i.e. reached in another
eigenspace of $T_{n\rightarrow 0}$.
In the following, we shall therefore analyse the transfer matrix
blocks corresponding to the non-trivial representations $D_\rho,\,
\rho \geq 1$. Each has $n+1-2\rho$ different eigenvalues of degeneracy
${n\choose \rho}-{n\choose \rho-1}$. By ordering the basis vectors
according to their permutation properties one can achieve a further
block-diagonalization of these blocks into ${n\choose \rho}-{n\choose
\rho-1}$ identical blocks of size $(n+1-2\rho)\times (n+1-2\rho)$ each
one containing every eigenvalue of the $D_\rho$-block exactly once. 
The transfer matrix blocks read
\begin{equation}
T_n^{(\rho)} (\sigma,\tau) = \ll ( 2 \sinh 2\beta J
)^\rho \, T_{n-2\rho}^{(0)} (\sigma-\rho,\tau-\rho;J,h) \gg \,,
\end{equation}
$\rho \leq \sigma,\tau \leq n-\rho$, where $T_n^{(0)}(\sigma,\tau;J,h)$ is
given by (\ref{trs}) without averaging over the quenched disorder.
In the limit $n\rightarrow 0$ we obtain for every positive number $\rho$
a different eigenvalue equation
\begin{equation}
\lambda^{(\rho)}\ \Phi ^{(\rho)}(x) =  \int_0^\infty dy
\ll \delta(x-f(y))\; (f'(y))^\rho \gg \ \Phi ^{(\rho)}(y) \
\label{eqvalrho}
\end{equation}
where we again used the function $f(x)$ defined in (\ref{f}). For every
eigenfunction $\Phi^{(1)}(x)$ of $T_{n\rightarrow 0}^{(1)}$ the
function $\frac{d}{dx}\Phi^{(1)}(x)$ is an eigenfunction of the RS
transfer matrix with the same eigenvalue. Only the largest eigenvalue
(equal to one) of $T_{n\rightarrow 0}^{(0)}$, corresponding to the 
density (\ref{density}),
cannot be reached by this procedure. Therefore the largest eigenvalue of
$T_{n\rightarrow 0}^{(1)}$ equals the second of $T_{n\rightarrow0}^{(0)}$ 
and so on. 

To unveil the physical meaning of the maximal eigenvalues of all
irreducible representations $\rho \geq 1$, one has to consider the
disorder--averaged moments of the spin correlation functions $\chi
^{(\rho )} _{ij} = \ll (\langle S_i S_j \rangle - \langle S_i
\rangle\langle S_j \rangle)^\rho \gg$. More precisely, we shall see
that they are dominated by
\begin{equation}
\chi ^{(\rho )} _{ij}\ \propto \ \left( \lambda^{(\rho)} \right)
^{|i-j|} \ , \label{correla}
\end{equation}
up to a multiplicative factor which depends on $\rho$.  To show this,
we consider the operator $\Sigma^{(\rho)}=(\Sigma\otimes{\bf 1}-{\bf
1}\otimes\Sigma) ^{\otimes\rho}$ where $\Sigma$ is the usual spin
operator~: $\Sigma | S \rangle = S|S\rangle $ .  The operator
$\Sigma^{(\rho)}$ is defined on a set of at least $2\rho$ identical
real replicas of the original disordered system. One can realize
that it maps the spaces carrying $D_0$ onto those carrying $D_\rho$. 
Since the ground state was found to be
replica symmetric for any kind of disorder and is a priori non
orthogonal to $\Sigma^{(\rho)} (D_\rho)$, the correlations of two
$\Sigma^{(\rho)}$ situated at sites $i$ and $j$ of the replicated
Ising chain decay exponentially as $(\lambda^{(\rho)})^{|j-i|}$. These
correlations are in addition equal to the left--hand side of
(\ref{correla}) as follows from the identity
\begin{equation}
\chi ^{(\rho )} _{ij} \ = \frac{1}{2^\rho}
\lim_{n\rightarrow 0} \hbox{ Trace} \left[
\Sigma^{(\rho)}_i (T_n) ^{j-i} \Sigma^{(\rho)}_j (T_n) ^{N-j} \right] \ .
\end{equation}
Therefore, the criterion
$\lambda^{(\rho)} \rightarrow 1$ for replica symmetry breaking
is equivalent to the divergence of the
correlation length $-(\log\,\lambda^{(\rho)} )^{-1}$ 
of the $\rho^{th}$ moment of the connected two-point
correlation function.  The case $\rho =2$ would correspond to the divergence
of the spin glass susceptibility $\chi ^{(2)} = \frac{1}{N} \sum_{i,j} 
( \langle S_i S_j \rangle - \langle S_i \rangle\langle S_j \rangle)^2$, 
whereas the linear susceptibility $\chi ^{(1)} = 
\frac{1}{N} \sum_{i,j} ( \langle S_i S_j \rangle - \langle S_i$
$ \rangle\langle S_j \rangle)$ remains finite. This behaviour can be
found in spin glass experiments as well as in mean-field theory 
\cite{review}.

As a result of a replica calculation, the validity of 
(\ref{eqvalrho},\ref{correla}) is questionable. It is a simple exercise
to check its correctness for the spin--glass chain without external fields~:
$\Phi ^{(\rho )} (x)=\delta (x -1)$ and $\lambda ^{(\rho )}
= \ll (\tanh (\beta J) )^\rho \gg$, for any $\rho >0$, in agreement with
known results \cite{Luck}. For a more generic distribution of the disorder,
exact calculations of the $\chi ^{(\rho )}$'s are much harder and, to our
knowledge, have been performed for $\rho =1$ only \cite{Luck}. 
We show now that our replica calculation is exact in this case.
Reproducing the approach of \cite{Neuw}, we compute the Fourier transform
of $\chi ^{(1)}$ by adding a small sinusoidal field of wave-number $q$
in the Hamiltonian (\ref{ham})
and expanding the free--energy at second order in the field to obtain
\begin{equation}
\chi  ^{(1)} (q) = \int dx\; da\; d\tilde a \; du\ \Phi _q
( x,a,\tilde a ,u ) \ \ll 2 u \left( \frac{ e^{-\beta J}}{ e^{\beta J} + 
x e^{-\beta J}} \right) - (a^2 + \tilde a^2) 
\left( \frac{ e^{-\beta J}}{ e^{\beta J} + x e^{-\beta J}} \right) ^2 \gg
\label{susc}
\end{equation}
where the invariant measure $\Phi _q$ fulfills
\begin{eqnarray}
\Phi _q( x,a,\tilde a ,u ) = \int dy\; db\; d\tilde b \; dv\ &&\Phi _q
( y,b,\tilde b ,v ) \ \ll \delta ( x - f(y))\; \delta ( a -2x -
( b \cos q + \tilde b \sin q ) f'(y) )  \nonumber \\
&& \delta ( \tilde a - (\tilde b \cos q - b \sin q) f'(y) )
\; \delta ( u - 2a+2x - v f'(y) -\frac 12 ( b^2 +\tilde b ^2) f''(y) )\gg.
\label{autoeq}
\end{eqnarray}
The susceptibility (\ref{susc}) depends on $\Phi _q$ through the average
values $[ u ]_q(x)$, $[a^2]_q(x)$ and $[\tilde a ^2]_q (x)$ only, where 
$[.]_q (x) = \int da d\tilde a du \Phi _q( x,a,\tilde a ,u ) (.)$ (note
that $[1]_q (x) = \Phi (x)$, see (\ref{density})). These three functions
obey three self--consistent equations obtained from (\ref{autoeq}) whose
integral kernels are regular. But their calculation still contains
the averages $[ a]_q (x)$ and $[\tilde a]_q (x)$ fulfilling
the complex equation
\begin{equation}
\int _0 ^{\infty} dy \; \left( \delta (x-y) - e^{-i q} \ll
\delta (x- f(y)). f'(y) \gg \right)
\ \big( [ a]_q (y) + i [\tilde a]_q (y) \big) =
2 x \Phi (x) \; ,
\label{fin}
\end{equation}
cf. (\ref{autoeq}). It becomes now clear that the pole $Q$ of 
$\chi ^{(1)}(q)$, closest to the real axis, is reached when 
$e^{i Q}= \lambda^{(1)}$, see (\ref{eqvalrho}). 
The asymptotic scaling (\ref{correla}) coincides 
therefore with exactly known results for $\chi ^{(1)}$. It would be
of high interest to extend the method of \cite{Neuw} (if possible) 
to test our replica predictions for larger values of $\rho \ge 2$. 

In this letter, we have studied the replica structure of one--dimensional 
Ising systems. From a technical point of view, we have first
block--diagonalized the transfer matrix according to the irreducible
representations of the permutation group of (finite) $n$ elements and
then sent $n$ to zero in the eigenvalue equation for each
representation.  This procedure has allowed us to understand the
physical meaning of the non trivial, ``replica symmmetry broken''
representations in terms of the moments of spin--spin correlations.
Though one--dimensional Ising models do not display any replica
symmetry breaking transition, we hope that the present approach will
be of interest in future to understand if such a phase transition
could occur in more realistic two or three--dimensional systems.

{\bf Acknowledgements:} We are grateful to D. Carpentier, A. Engel,
J. Kurchan, and D. Malzahn for interesting and helpful discussions.
M.W. thanks the Laboratoire de Physique Th\'eorique de
l'ENS for its hospitality.


\begin{thebibliography}{[99]}
\bibitem{review} K. Binder, A.P. Young, {\it Rev. Mod. Phys.} {\bf 58},
801 (1986)
\bibitem{SK} D. Sherrington, S. Kirkpatrick, {\it Phys. Rev. Lett.} {\bf 64},
 1972 (1975)
\bibitem{Par1} G. Parisi, {\it Phys. Lett.} {\bf 73A},  203 (1979); 
{\it J. Phys. A} {\bf 13}, 1101 (1980)
\bibitem{mezard} M.M\'ezard {\it et al., J. Physique} {\bf 45}, 843 (1984) \\
M. M\'ezard, M. A. Virasoro {\it J. Physique} {\bf 46}, 1293 (1985)
\bibitem{fisher} D. Huse, D. Fisher, {\it J. Phys. A} {\bf 20}, L997 (1987)
\bibitem{Luck} J.M. Luck, {\it Syst\`emes d\'esordonn\'es unidimensionels} 
(Al\'ea Saclay, 1992)
\bibitem{pendry} J.B. Pendry, {\it J. Phys. C} {\bf 15}, 4821 (1982)
\bibitem{Derr2} B. Derrida, J. Vannimenus, Y. Pomeau, {\it J. Phys.C}
{\bf 11}, 4749 (1978) 
\bibitem{Luck2}J.M. Luck, M. Funke, Th.M. Nieuwenhuizen, {\it J.Phys.A}
{\bf 24}, 4155 (1991)
\bibitem{bray} A.J. Bray, M.A. Moore, {\it J. Phys. A} {\bf 18}, L683 (1985)
\bibitem{Wig} E.P. Wigner, {\it Group theory and its applications to the
quantum mechanics of atomic spectra} (Academic Press, New York, 1959)
\bibitem{Derr} B. Derrida, H. J. Hilhorst, {\it J. Phys. A}
{\bf 16}, 2641 (1983) 
\bibitem{Lin} T.F. Lin, {\it J. Math. Phys.} {\bf 11}, 1584 (1970)
\bibitem{Neuw} J.M. Luck, Th.M. Nieuwenhuizen, {\it J. Phys. A} 
{\bf 22}, 2151 (1989)
\end{thebibliography}
\end{document}